\documentclass[10pt,letterpaper]{article}
\usepackage[top=0.85in,left=2.75in,footskip=0.75in,marginparwidth=2in]{geometry}

\usepackage{cite}

\usepackage{nameref,hyperref}
\usepackage[utf8x]{inputenc}
\usepackage[right]{lineno}

\usepackage{fontawesome}
\usepackage{xcolor}
\usepackage[framemethod=tikz]{mdframed}
\usepackage{lipsum}

\mdfdefinestyle{mystyle}{%
linecolor=green!40!black,outerlinewidth=1pt,%
frametitlerule=true,frametitlefont=\sffamily\bfseries\color{white},%
frametitlerulewidth=1pt,frametitlerulecolor=green!40!black,%
frametitlebackgroundcolor=green!40!black,
backgroundcolor=green!5,
innertopmargin=\topskip,
roundcorner=5pt
}
\newmdenv[style=mystyle]{exa}
\newenvironment{example}[1]
  {\begin{exa}[frametitle=#1]}
  {\end{exa}}

\usepackage{microtype}
\DisableLigatures[f]{encoding = *, family = * }

\raggedright
\usepackage{changepage}

\usepackage[aboveskip=1pt,labelfont=bf,labelsep=period,singlelinecheck=off]{caption}

\makeatletter
\renewcommand{\@biblabel}[1]{\quad#1.}
\makeatother

\usepackage{lastpage,fancyhdr,graphicx}
\usepackage{epstopdf}
\fancyheadoffset[L]{2.25in}
\fancyfootoffset[L]{2.25in}

\usepackage{color}

\definecolor{Gray}{gray}{.25}

\usepackage{graphicx}

\usepackage{sidecap}

\usepackage{wrapfig}
\usepackage[pscoord]{eso-pic}
\usepackage[fulladjust]{marginnote}
\reversemarginpar

\begin{document}
\vspace*{0.35in}

\begin{flushleft}
{\Large
\textbf\newline{Super-resolution structured illumination microscopy: past, present and future}
}
\newline
\\
Kirti Prakash\textsuperscript{1,2,*, },
Benedict Diederich\textsuperscript{3,4},
Stefanie Reichelt\textsuperscript{5},
Rainer Heintzmann\textsuperscript{3,4,6,*},
Lothar Schermelleh\textsuperscript{7,*}
\\
\bigskip
\bf{1} National Physical Laboratory, TW11 0LW Teddington, UK
\\
\bf{2} Department of Paediatrics, Wellcome–MRC Cambridge Stem Cell Institute, University of Cambridge, Cambridge, UK
\\
\bf{3} Leibniz Institute of Photonic Technology, Albert-Einstein-Straße 9, 07745 Jena, Germany
\\
\bf{4} Institute of Physical Chemistry and Abbe Center of Photonics, Helmholtzweg 4, Friedrich-Schiller-University, Jena, Germany
\\
\bf{5} CRUK Cambridge Research Institute, Robinson Way, Cambridge CB2 0RE, UK
\\
\bf{6} Faculty of Physics and Astronomy, Friedrich-Schiller-University, Jena, Germany
\\
\bf{7} Micron Advanced Bioimaging Unit, Department of Biochemistry, University of Oxford, Oxford OX1 3QU, UK
\\
\bf{\faTwitter} \href{https://twitter.com/kirtiprakash25}{kirtiprakash25}, \href{https://twitter.com/beniroquai}{beniroquai}, \href{https://twitter.com/StefanieReiche6}{StefanieReiche6}, \href{https://twitter.com/HeintzmannLab}{HeintzmannLab}, \href{https://twitter.com/LSchermelleh}{LSchermelleh}
\\
\bigskip
* heintzmann@gmail.com, kirtiprakash2.71@gmail.com, lothar.schermelleh@bioch.ox.ac.uk

%

\end{flushleft}

\section*{Abstract}
Structured illumination microscopy (SIM) has emerged as an essential technique for 3D and live-cell super-resolution imaging. However, to date, there has not been a dedicated workshop or journal issue covering the various aspects of SIM, from bespoke hardware and software development and the use of commercial instruments to biological applications. This special issue aims to recap recent developments as well as outline future trends. In addition to SIM, we cover related topics such as complementary super-resolution microscopy techniques, computational imaging, visualisation and image processing methods. 

\section*{Introduction}
Fluorescence light microscopy is a core technique in life sciences that has contributed to countless major discoveries. However, the diffraction limit, first described by Ernst Abbe in 1873 \cite{abbe1873}, has restricted the optical resolution to about half the wavelength of the light used, i.e.  200-300 nm in the lateral directions (x and y), and to about the wavelength of light, i.e. 500-800 nm along the optical axis (z). In the past two decades, several super-resolution microscopy (SRM) approaches have been developed that allow to overcome this barrier and push the spatial resolution to the 10-100 nm range, thus closing the gap to electron microscopy \cite{schermelleh2010guide, heintzmann2017super, schermelleh2019super}. The SRM techniques can be further divided into single-molecule localisation microscopy (SMLM) \cite{lidke2005superresolution}, that includes techniques like photoactivatable localisation microscopy (PALM) \cite{betzig2006imaging, hess2006ultra} and stochastic optical reconstruction microscopy (STORM) \cite{rust2006stochastic}, stimulated emission depletion (STED) \cite{Okhonin1987,hell1994breaking,baer1999method} microscopy and structured illumination microscopy (SIM) \cite{heintzmann1999laterally, gustafsson2000surpassing}. In recognition of these breakthrough developments Eric Betzig, Stefan Hell and William E. Moerner were awarded the Nobel prize for Chemistry in 2014 \cite{betzig2014nobel}.

\begin{example}{Frequently asked questions in super-resolution structured illumination microscopy field}
The following questions have not always been agreed upon in the super-resolution/SIM field, and with this special issue, we hope to address some of these:
\begin{enumerate}
\item 
What is super-resolution microscopy and should diffraction-limited linear SIM be classified as `super-resolution'?
\item
Should High-NA TIRF-SIM, which can achieve lateral resolution down to 84 nm, be considered as diffraction limited?
\item
Can non-linear SIM  become broadly applicable and live-cell compatible?
\item
Do you need `switching' of states for non-linear super-resolution?
\item
How can information about single-molecule detection be best combined with the knowledge of the illumination structure?
\item
Do high quality SIM images require reconstruction in Fourier space?
\item
Can SIM be used for deep tissue imaging?
\item
How can the fundamental limitation of SIM, i.e. generating sufficient stripe contrast in densely labelled and/or extended biological structures due to out-of-focus light, be addressed?
\item
Should image scanning microscopy be considered a form of SIM and what forms of structured illumination could be used other than stripes?
\item
Can SIM be used to improve the resolution of (Rayleigh scattering) transmission microscopy?
\item
How does sparse illumination compare to dense illumination in linear and non-linear SIM?
\item
Can we generate “true” super-resolution images from simple instruments enhanced with machine-learning-based algorithms?
\item
Can research-grade super-resolution (SIM) microscopes be built cost-efficiently?
\end{enumerate}
\end{example}

Recent SRM developments have focussed on combining localisation-based microscopy with modulated illumination to push the resolution (precision) towards a few nanometers as in MINFLUX and SIMFLUX \cite{balzarotti2017nanometer, cnossen2020localization, reymond2020modulation, prakash2021molecular}. However, an increase in resolution requires an increase in local light dosage, which increases photobleaching/phototoxicity (SIM, STED, SMLM) or requires the need to introduce spatial and temporal sparsity (MINFLUX) making imaging comparably slow. Moreover, most SRM techniques mentioned above rely primarily on chemically fixed (i.e. dead) cells, are restricted in the imaged sample volume (2D, single plane, small field-of-views) or are restricted to imaging sparse single entities such as vesicles or single molecules.

\begin{figure}[b!]
\begin{center}
\includegraphics[width=\linewidth]{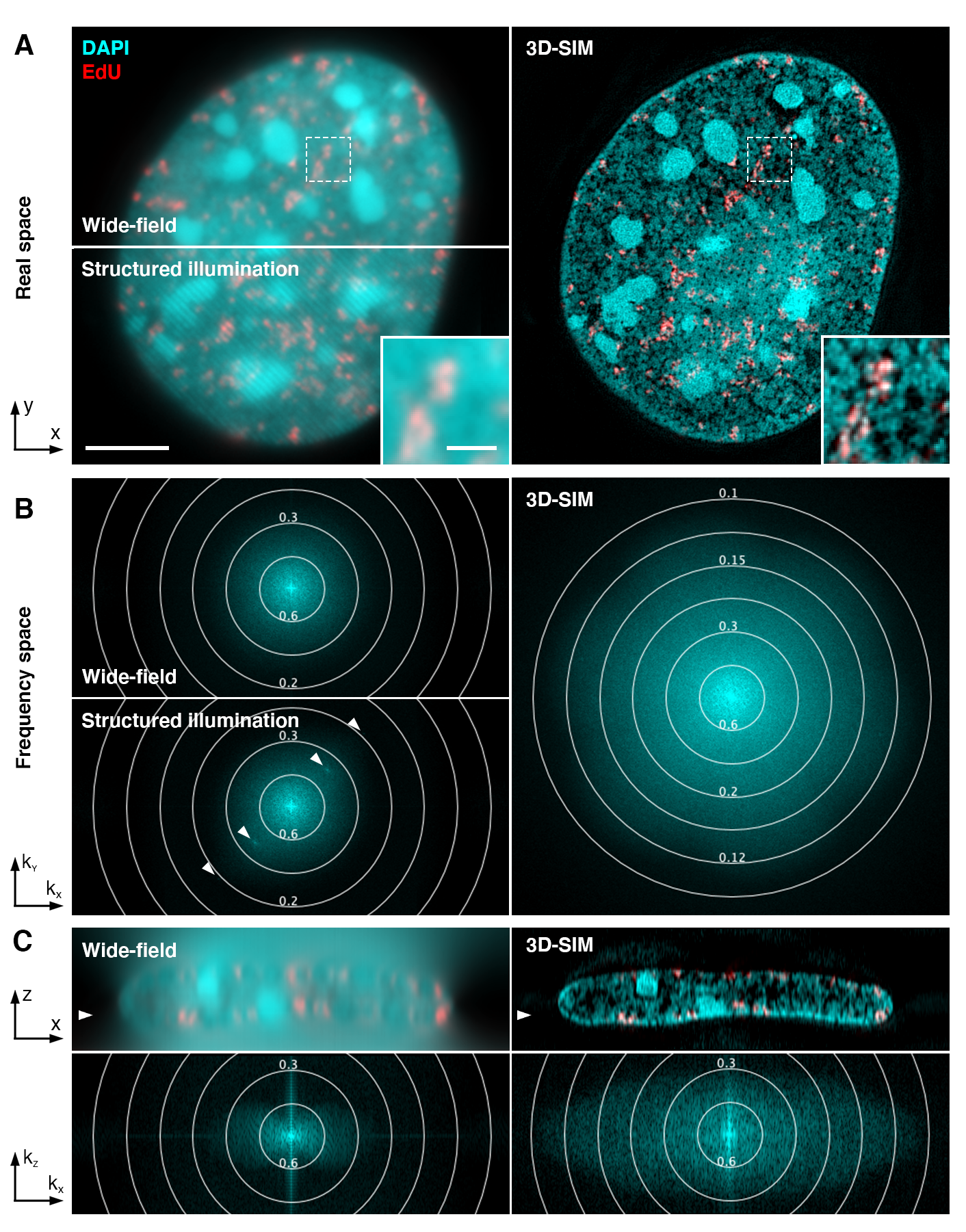}
\end{center}
\caption[]{(Caption next page.)} \label{fig:sim}
\end{figure}

Super-resolution linear SIM is a notable exception \cite{heintzmann2009subdiffraction, o2014optimized}. By making use of frequency mixing when exciting samples with a patterned illumination followed by computational unmixing and reconstruction, SIM achieves a 2-fold resolution increase over conventional diffraction-limited fluorescence microscopy in 2D or 3D (Figure~\ref{fig:sim}). While the numerical resolution improvement is moderate compared to other SRM techniques, it pushes SIM into an application sweet spot with most macromolecular structures and their dynamics falling in the size range of 100-300 nm \cite{schermelleh2008subdiffraction, smeets2014three, shaw2017creim}. By not asking for the highest spatial resolution, SIM is less demanding in terms of photon budget and therefore more compatible with live-cell imaging \cite{kner2009super, hirvonen2009structured, shao2011super,mudry2012structured, huang2018fast, wu2018faster}. In addition, it offers volumetric imaging with relatively large field-of-views and comparably high temporal resolution which makes it suitable for high-content and live-cell imaging \cite{york2012resolution,york2013instant, mertz2011optical}. Taken together, the uniquely balanced combination of properties have made SIM remarkably successful in a wide range of biological applications and promoting new discoveries  \cite{gray2019nanoscale, nozumi2017coordinated, jacquemet2019filopodome, brown2011remodelling, mcarthur2018bak, eswaramoorthy2014asymmetric, miron2020chromatin, murugesan2016formin, hubner2015remodeling, sir2011primary, manor2015mitochondria, guo2018visualizing, lesterlin2014reca, ochs2019stabilization, smeets2014three, sonnen20123d, mennella2012subdiffraction}.

\section*{Recent and future developments in SIM}
SIM has been available in commercial instruments for more than 10 years and has been successfully established in many labs and core facilities around the world. However, wider dissemination of SIM has been curtailed by the complexity of instruments and its proneness to reconstruction artefacts when sample properties, system calibration and parameter settings are not carefully matched \cite{demmerle2017strategic}. Thus, to lower the activation energy for research labs to venture into super-resolution SIM, many recent developments aim at further performance/resolution enhancement and `democratising’ SIM by
\begin{itemize}
\item 
using non-linear SIM approaches \cite{heintzmann2002saturated, gustafsson2005nonlinear, rego2012nonlinear, li2015extended} or by combinations with single-molecule imaging \cite{balzarotti2017nanometer, reymond2019simple, gu2019molecular, cnossen2020localization,jouchet2019nanometric}
\item
exploiting correlative and combinatorial fluorescence imaging approaches with multifocal \cite{hajj2014whole, abrahamsson2017multifocus}, 2-photon \cite{ingaramo2014two, sheppard2017image, liu2021two} and light-sheet microscopy \cite{keller2010fast, chang2017csilsfm}
\item 
implementing the technique into smaller and more cost-efficient setups \cite{helle2020structured, wang2021ucsim2, lachetta2020simulating}
\item
making the method more robust against artefacts using alternative illumination schemes \cite{schaefer2004structured, ball2015simcheck, mo2021structured} and/or intelligent data processing \cite{yeh2019computational, boland2020improving}
\item
increasing its application range e.g. by the implementation of adaptive optics \cite{debarre2008adaptive, booth2015aberrations, ji2017adaptive}, cryo-imaging \cite{phillips2020cryosim, sonnen20123d}
\end{itemize}
Important advancements are being made to improve and simplify the instrumentation. For instance, the newly designed optical setups using an open-sources toolbox \cite{wang2021ucsim2} or waveguide-based photonic chips \cite{yeh2019computational} or open-source software development \cite{muller2016open} may reduce the cost of SR imaging enabling SIM on smaller and more affordable devices. Moreover, next-generation sCMOS cameras (for example, the new 500 fps sCMOS by Photometrics) offer increased frame rates and higher sensitivity with reduced noise levels to enable faster high-quality SIM imaging for observing cellular dynamics \cite{chen2014lattice, liu2018observing, shaw2015high}. For a maximal imaging speed, the rolling shutter schemes can be combined with SIM data acquisition \cite{song2016fast}.\\
In parallel, the rapidly evolving field of computational microscopy is paving the way to circumvent the resolution limit without the necessity of specialised hardware.  Such computational techniques not only reduce the complexity of the instrumentation but can also potentially reduce aberration artefacts \cite{yeh2019computational, boland2020improving, Prakash121061, wang2021ucsim2, lachetta2020simulating, russell2021mmsim}. Adaptive optics (AO) works well to correct sample aberrations. However, it does not solve the problem of reduced illumination pattern contrast in the presence of significant out of focus background. One solution here is offered by combining SIM with light sheet illumination, 2-photon, or photoswitching methods.   \\
Recent advancements in intelligent microscopy and machine learning (ML) assisted denoising are now being used for improving SIM reconstructions. For example, using deep-learning models, the apparent axial resolution of SIM has been improved by a factor of two \cite{jin2020deep, ling2020fast, boland2020improving}. By combining physical data acquisition with simple image processing algorithms, one recover previously inaccessible information. At the same time, this brings up new problems of validation of the results from ML methods.  New standards have to be developed, where computational scientists and mathematicians can help to bring together state-of-the-art algorithms, such as GPU-parallelization for faster data processing or machine-learning-based data interpretation.\\

\section*{The wider-reaching social implications}

SRM has evolved into a highly interdisciplinary field requiring experts from physics, engineering, chemistry, biology and computer sciences to drive innovations for increasing spatial and temporal resolution and ultimately biological applicability.  The global advances in super-resolution microscopy need to be synchronized to inspire researchers in the field of microscopy to combine their methods with techniques from different areas. 

SRM and SIM, in particular, have become important tools of basic science discovery. However, there are still considerable activation barriers for the researcher to use SRM, as instruments are not yet `turn-key', and using SIM (or any other SRM method) to its full potential requires considerable knowledge. Furthermore, the use of SRM to study human pathologies and for medical diagnosis is still in its infancy, but will foreseeably play a considerable role in the future. New SIM modalities, implementation of AO, machine-learning, and the intelligent algorithms will help to democratise SIM and increase it's usage in basic research and biomedical application.

Super-resolution microscopy is very often an expensive undertaking, thus limiting its broader application. In this meeting, we plan to kick-off the idea of an open-source “openSIM” system inspired by its derivatives in the light-sheet community. Open discussions where ideas and requirements for a prototype are gathered are a first and significant step to make science not only affordable but available. This special issue shows current progress in SIM development and application.

\section*{Acknowledgements}
The special issue is part of Theo Murphy international scientific meeting organised by the Royal Society called \href{https://royalsociety.org/science-events-and-lectures/2020/05/SIMposium/}{SIMposium: recent advancements in structured illumination microscopy.}
We thank Mike Shaw for comments on the manuscript.  A Twitter (\faTwitter) discussion on this topic which can be followed \href{https://twitter.com/kirtiprakash25/status/1152500999606755334}{here} or with hashtag SIMposium.

\nolinenumbers

\bibliography{mylib}

\begin{thebibliography}{10}

\bibitem{abbe1873}
E.~Abbe.
\newblock Beitrage zur {T}heorie des {M}ikroskops und der mikroskopischen
  {W}ahrnehmung.
\newblock {\em Archiv für Mikroskopische Anatomie}, 9:413--420, 1873.

\bibitem{abrahamsson2017multifocus}
S.~Abrahamsson, H.~Blom, A.~Agostinho, D.~C. Jans, A.~Jost, M.~M{\"u}ller,
  L.~Nilsson, K.~Bernhem, T.~J. Lambert, R.~Heintzmann, et~al.
\newblock Multifocus structured illumination microscopy for fast volumetric
  super-resolution imaging.
\newblock {\em Biomedical optics express}, 8(9):4135--4140, 2017.

\bibitem{baer1999method}
S.~C. Baer.
\newblock Method and apparatus for improving resolution in scanned optical
  system, Feb.~2 1999.
\newblock US Patent 5,866,911.

\bibitem{ball2015simcheck}
G.~Ball, J.~Demmerle, R.~Kaufmann, I.~Davis, I.~M. Dobbie, and L.~Schermelleh.
\newblock Simcheck: a toolbox for successful super-resolution structured
  illumination microscopy.
\newblock {\em Scientific reports}, 5(1):1--12, 2015.

\bibitem{balzarotti2017nanometer}
F.~Balzarotti, Y.~Eilers, K.~C. Gwosch, A.~H. Gynn{\aa}, V.~Westphal, F.~D.
  Stefani, J.~Elf, and S.~W. Hell.
\newblock Nanometer resolution imaging and tracking of fluorescent molecules
  with minimal photon fluxes.
\newblock {\em Science}, 355(6325):606--612, 2017.

\bibitem{betzig2014nobel}
E.~Betzig, S.~W. Hell, and W.~E. Moerner.
\newblock The nobel prize in chemistry 2014.
\newblock {\em Nobel Media AB}, 2014.

\bibitem{betzig2006imaging}
E.~Betzig, G.~H. Patterson, R.~Sougrat, O.~W. Lindwasser, S.~Olenych, J.~S.
  Bonifacino, M.~W. Davidson, J.~Lippincott-Schwartz, and H.~F. Hess.
\newblock Imaging intracellular fluorescent proteins at nanometer resolution.
\newblock {\em Science}, 313(5793):1642--1645, 2006.

\bibitem{boland2020improving}
M.~Boland, E.~A. Cohen, S.~Flaxman, and M.~A. Neil.
\newblock Improving axial resolution in sim using deep learning.
\newblock {\em arXiv preprint arXiv:2009.02264}, 2020.

\bibitem{booth2015aberrations}
M.~Booth, D.~Andrade, D.~Burke, B.~Patton, and M.~Zurauskas.
\newblock Aberrations and adaptive optics in super-resolution microscopy.
\newblock {\em Microscopy}, 64(4):251--261, 2015.

\bibitem{brown2011remodelling}
A.~C. Brown, S.~Oddos, I.~M. Dobbie, J.-M. Alakoskela, R.~M. Parton,
  P.~Eissmann, M.~A. Neil, C.~Dunsby, P.~M. French, I.~Davis, et~al.
\newblock Remodelling of cortical actin where lytic granules dock at natural
  killer cell immune synapses revealed by super-resolution microscopy.
\newblock {\em PLoS Biol}, 9(9):e1001152, 2011.

\bibitem{chang2017csilsfm}
B.-J. Chang, V.~D.~P. Meza, and E.~H. Stelzer.
\newblock csilsfm combines light-sheet fluorescence microscopy and coherent
  structured illumination for a lateral resolution below 100 nm.
\newblock {\em Proceedings of the National Academy of Sciences},
  114(19):4869--4874, 2017.

\bibitem{chen2014lattice}
B.-C. Chen, W.~R. Legant, K.~Wang, L.~Shao, D.~E. Milkie, M.~W. Davidson,
  C.~Janetopoulos, X.~S. Wu, J.~A. Hammer, Z.~Liu, et~al.
\newblock Lattice light-sheet microscopy: imaging molecules to embryos at high
  spatiotemporal resolution.
\newblock {\em Science}, 346(6208), 2014.

\bibitem{cnossen2020localization}
J.~Cnossen, T.~Hinsdale, R.~{\O}. Thorsen, M.~Siemons, F.~Schueder,
  R.~Jungmann, C.~S. Smith, B.~Rieger, and S.~Stallinga.
\newblock Localization microscopy at doubled precision with patterned
  illumination.
\newblock {\em Nature methods}, 17(1):59--63, 2020.

\bibitem{debarre2008adaptive}
D.~D{\'e}barre, E.~J. Botcherby, M.~J. Booth, and T.~Wilson.
\newblock Adaptive optics for structured illumination microscopy.
\newblock {\em Optics express}, 16(13):9290--9305, 2008.

\bibitem{demmerle2017strategic}
J.~Demmerle, C.~Innocent, A.~J. North, G.~Ball, M.~M{\"u}ller, E.~Miron,
  A.~Matsuda, I.~M. Dobbie, Y.~Markaki, and L.~Schermelleh.
\newblock Strategic and practical guidelines for successful structured
  illumination microscopy.
\newblock {\em Nature protocols}, 12(5):988--1010, 2017.

\bibitem{eswaramoorthy2014asymmetric}
P.~Eswaramoorthy, P.~W. Winter, P.~Wawrzusin, A.~G. York, H.~Shroff, and K.~S.
  Ramamurthi.
\newblock Asymmetric division and differential gene expression during a
  bacterial developmental program requires diviva.
\newblock {\em PLoS Genet}, 10(8):e1004526, 2014.

\bibitem{gray2019nanoscale}
R.~D. Gray, D.~Albrecht, C.~Beerli, M.~Huttunen, G.~H. Cohen, I.~J. White,
  J.~J. Burden, R.~Henriques, and J.~Mercer.
\newblock Nanoscale polarization of the entry fusion complex of vaccinia virus
  drives efficient fusion.
\newblock {\em Nature microbiology}, 4(10):1636--1644, 2019.

\bibitem{gu2019molecular}
L.~Gu, Y.~Li, S.~Zhang, Y.~Xue, W.~Li, D.~Li, T.~Xu, and W.~Ji.
\newblock Molecular resolution imaging by repetitive optical selective
  exposure.
\newblock {\em Nature methods}, 16(11):1114--1118, 2019.

\bibitem{guo2018visualizing}
Y.~Guo, D.~Li, S.~Zhang, Y.~Yang, J.-J. Liu, X.~Wang, C.~Liu, D.~E. Milkie,
  R.~P. Moore, U.~S. Tulu, et~al.
\newblock Visualizing intracellular organelle and cytoskeletal interactions at
  nanoscale resolution on millisecond timescales.
\newblock {\em Cell}, 175(5):1430--1442, 2018.

\bibitem{gustafsson2000surpassing}
M.~G. Gustafsson.
\newblock Surpassing the lateral resolution limit by a factor of two using
  structured illumination microscopy.
\newblock {\em Journal of microscopy}, 198(2):82--87, 2000.

\bibitem{gustafsson2005nonlinear}
M.~G. Gustafsson.
\newblock Nonlinear structured-illumination microscopy: wide-field fluorescence
  imaging with theoretically unlimited resolution.
\newblock {\em Proceedings of the National Academy of Sciences},
  102(37):13081--13086, 2005.

\bibitem{hajj2014whole}
B.~Hajj, J.~Wisniewski, M.~El~Beheiry, J.~Chen, A.~Revyakin, C.~Wu, and
  M.~Dahan.
\newblock Whole-cell, multicolor superresolution imaging using volumetric
  multifocus microscopy.
\newblock {\em Proceedings of the National Academy of Sciences},
  111(49):17480--17485, 2014.

\bibitem{heintzmann1999laterally}
R.~Heintzmann and C.~G. Cremer.
\newblock Laterally modulated excitation microscopy: improvement of resolution
  by using a diffraction grating.
\newblock In {\em Optical Biopsies and Microscopic Techniques III}, volume
  3568, pages 185--196. International Society for Optics and Photonics, 1999.

\bibitem{heintzmann2009subdiffraction}
R.~Heintzmann and M.~G. Gustafsson.
\newblock Subdiffraction resolution in continuous samples.
\newblock {\em Nature Photonics}, 3(7):362--364, 2009.

\bibitem{heintzmann2017super}
R.~Heintzmann and T.~Huser.
\newblock Super-resolution structured illumination microscopy.
\newblock {\em Chemical reviews}, 117(23):13890--13908, 2017.

\bibitem{heintzmann2002saturated}
R.~Heintzmann, T.~M. Jovin, and C.~Cremer.
\newblock Saturated patterned excitation microscopy—a concept for optical
  resolution improvement.
\newblock {\em JOSA A}, 19(8):1599--1609, 2002.

\bibitem{hell1994breaking}
S.~W. Hell and J.~Wichmann.
\newblock Breaking the diffraction resolution limit by stimulated emission:
  stimulated-emission-depletion fluorescence microscopy.
\newblock {\em Optics letters}, 19(11):780--782, 1994.

\bibitem{helle2020structured}
{\O}.~I. Helle, F.~T. Dullo, M.~Lahrberg, J.-C. Tinguely, O.~G. Helles{\o}, and
  B.~S. Ahluwalia.
\newblock Structured illumination microscopy using a photonic chip.
\newblock {\em Nature Photonics}, 14(7):431--438, 2020.

\bibitem{hess2006ultra}
S.~T. Hess, T.~P. Girirajan, and M.~D. Mason.
\newblock Ultra-high resolution imaging by fluorescence photoactivation
  localization microscopy.
\newblock {\em Biophysical journal}, 91(11):4258, 2006.

\bibitem{hirvonen2009structured}
L.~M. Hirvonen, K.~Wicker, O.~Mandula, and R.~Heintzmann.
\newblock Structured illumination microscopy of a living cell.
\newblock {\em European Biophysics Journal}, 38(6):807--812, 2009.

\bibitem{huang2018fast}
X.~Huang, J.~Fan, L.~Li, H.~Liu, R.~Wu, Y.~Wu, L.~Wei, H.~Mao, A.~Lal, P.~Xi,
  et~al.
\newblock Fast, long-term, super-resolution imaging with hessian structured
  illumination microscopy.
\newblock {\em Nature biotechnology}, 36(5):451--459, 2018.

\bibitem{hubner2015remodeling}
B.~H{\"u}bner, M.~Lomiento, F.~Mammoli, D.~Illner, Y.~Markaki, S.~Ferrari,
  M.~Cremer, and T.~Cremer.
\newblock Remodeling of nuclear landscapes during human myelopoietic cell
  differentiation maintains co-aligned active and inactive nuclear
  compartments.
\newblock {\em Epigenetics \& chromatin}, 8(1):1--21, 2015.

\bibitem{ingaramo2014two}
M.~Ingaramo, A.~G. York, P.~Wawrzusin, O.~Milberg, A.~Hong, R.~Weigert,
  H.~Shroff, and G.~H. Patterson.
\newblock Two-photon excitation improves multifocal structured illumination
  microscopy in thick scattering tissue.
\newblock {\em Proceedings of the National Academy of Sciences},
  111(14):5254--5259, 2014.

\bibitem{jacquemet2019filopodome}
G.~Jacquemet, A.~Stubb, R.~Saup, M.~Miihkinen, E.~Kremneva, H.~Hamidi, and
  J.~Ivaska.
\newblock Filopodome mapping identifies p130cas as a mechanosensitive regulator
  of filopodia stability.
\newblock {\em Current Biology}, 29(2):202--216, 2019.

\bibitem{ji2017adaptive}
N.~Ji.
\newblock Adaptive optical fluorescence microscopy.
\newblock {\em Nature methods}, 14(4):374--380, 2017.

\bibitem{jin2020deep}
L.~Jin, B.~Liu, F.~Zhao, S.~Hahn, B.~Dong, R.~Song, T.~C. Elston, Y.~Xu, and
  K.~M. Hahn.
\newblock Deep learning enables structured illumination microscopy with low
  light levels and enhanced speed.
\newblock {\em Nature communications}, 11(1):1--7, 2020.

\bibitem{jouchet2019nanometric}
P.~Jouchet, C.~Cabriel, N.~Bourg, M.~Bardou, C.~Po{\"u}s, E.~Fort, and
  S.~L{\'e}v{\^e}que-Fort.
\newblock Nanometric axial localization of single fluorescent molecules with
  modulated excitation.
\newblock {\em BioRxiv}, page 865865, 2019.

\bibitem{keller2010fast}
P.~J. Keller, A.~D. Schmidt, A.~Santella, K.~Khairy, Z.~Bao, J.~Wittbrodt, and
  E.~H. Stelzer.
\newblock Fast, high-contrast imaging of animal development with scanned light
  sheet--based structured-illumination microscopy.
\newblock {\em Nature methods}, 7(8):637--642, 2010.

\bibitem{kner2009super}
P.~Kner, B.~B. Chhun, E.~R. Griffis, L.~Winoto, and M.~G. Gustafsson.
\newblock Super-resolution video microscopy of live cells by structured
  illumination.
\newblock {\em Nature methods}, 6(5):339--342, 2009.

\bibitem{lachetta2020simulating}
M.~Lachetta, H.~Sandmeyer, A.~Sandmeyer, J.~S. am~Esch, T.~Huser, and
  M.~M{\"u}ller.
\newblock Simulating digital micromirror devices for patterning coherent
  excitation light in structured illumination microscopy.
\newblock {\em bioRxiv}, 2020.

\bibitem{lesterlin2014reca}
C.~Lesterlin, G.~Ball, L.~Schermelleh, and D.~J. Sherratt.
\newblock Reca bundles mediate homology pairing between distant sisters during
  dna break repair.
\newblock {\em Nature}, 506(7487):249--253, 2014.

\bibitem{li2015extended}
D.~Li, L.~Shao, B.-C. Chen, X.~Zhang, M.~Zhang, B.~Moses, D.~E. Milkie, J.~R.
  Beach, J.~A. Hammer, M.~Pasham, et~al.
\newblock Extended-resolution structured illumination imaging of endocytic and
  cytoskeletal dynamics.
\newblock {\em Science}, 349(6251), 2015.

\bibitem{lidke2005superresolution}
K.~A. Lidke, B.~Rieger, T.~M. Jovin, and R.~Heintzmann.
\newblock Superresolution by localization of quantum dots using blinking
  statistics.
\newblock {\em Optics express}, 13(18):7052--7062, 2005.

\bibitem{ling2020fast}
C.~Ling, C.~Zhang, M.~Wang, F.~Meng, L.~Du, and X.~Yuan.
\newblock Fast structured illumination microscopy via deep learning.
\newblock {\em Photonics Research}, 8(8):1350--1359, 2020.

\bibitem{liu2021two}
F.~Liu, Q.~Li, S.~Jiang, L.~Zhou, J.~Zhang, and H.~Zhang.
\newblock Two-photon structured illumination microscopy imaging using fourier
  ptychography scheme.
\newblock {\em Optics Communications}, page 126872, 2021.

\bibitem{liu2018observing}
T.-L. Liu, S.~Upadhyayula, D.~E. Milkie, V.~Singh, K.~Wang, I.~A. Swinburne,
  K.~R. Mosaliganti, Z.~M. Collins, T.~W. Hiscock, J.~Shea, et~al.
\newblock Observing the cell in its native state: Imaging subcellular dynamics
  in multicellular organisms.
\newblock {\em Science}, 360(6386), 2018.

\bibitem{manor2015mitochondria}
U.~Manor, S.~Bartholomew, G.~Golani, E.~Christenson, M.~Kozlov, H.~Higgs,
  J.~Spudich, and J.~Lippincott-Schwartz.
\newblock A mitochondria-anchored isoform of the actin-nucleating spire protein
  regulates mitochondrial division.
\newblock {\em Elife}, 4:e08828, 2015.

\bibitem{mcarthur2018bak}
K.~McArthur, L.~W. Whitehead, J.~M. Heddleston, L.~Li, B.~S. Padman,
  V.~Oorschot, N.~D. Geoghegan, S.~Chappaz, S.~Davidson, H.~San~Chin, et~al.
\newblock Bak/bax macropores facilitate mitochondrial herniation and mtdna
  efflux during apoptosis.
\newblock {\em Science}, 359(6378), 2018.

\bibitem{mennella2012subdiffraction}
V.~Mennella, B.~Keszthelyi, K.~McDonald, B.~Chhun, F.~Kan, G.~C. Rogers,
  B.~Huang, and D.~Agard.
\newblock Subdiffraction-resolution fluorescence microscopy reveals a domain of
  the centrosome critical for pericentriolar material organization.
\newblock {\em Nature cell biology}, 14(11):1159--1168, 2012.

\bibitem{mertz2011optical}
J.~Mertz.
\newblock Optical sectioning microscopy with planar or structured illumination.
\newblock {\em Nature methods}, 8(10):811, 2011.

\bibitem{miron2020chromatin}
E.~Miron, R.~Oldenkamp, J.~M. Brown, D.~M. Pinto, C.~S. Xu, A.~R. Faria, H.~A.
  Shaban, J.~D. Rhodes, C.~Innocent, S.~de~Ornellas, et~al.
\newblock Chromatin arranges in chains of mesoscale domains with nanoscale
  functional topography independent of cohesin.
\newblock {\em Science advances}, 6(39):eaba8811, 2020.

\bibitem{mo2021structured}
Y.~Mo, F.~Feng, H.~Mao, J.~Fan, and L.~Chen.
\newblock Structured illumination microscopy artifacts caused by illumination
  scattering.
\newblock {\em bioRxiv}, pages 2021--01, 2021.

\bibitem{mudry2012structured}
E.~Mudry, K.~Belkebir, J.~Girard, J.~Savatier, E.~Le~Moal, C.~Nicoletti,
  M.~Allain, and A.~Sentenac.
\newblock Structured illumination microscopy using unknown speckle patterns.
\newblock {\em Nature Photonics}, 6(5):312--315, 2012.

\bibitem{muller2016open}
M.~M{\"u}ller, V.~M{\"o}nkem{\"o}ller, S.~Hennig, W.~H{\"u}bner, and T.~Huser.
\newblock Open-source image reconstruction of super-resolution structured
  illumination microscopy data in imagej.
\newblock {\em Nature communications}, 7(1):1--6, 2016.

\bibitem{murugesan2016formin}
S.~Murugesan, J.~Hong, J.~Yi, D.~Li, J.~R. Beach, L.~Shao, J.~Meinhardt,
  G.~Madison, X.~Wu, E.~Betzig, et~al.
\newblock Formin-generated actomyosin arcs propel t cell receptor microcluster
  movement at the immune synapse.
\newblock {\em Journal of Cell Biology}, 215(3):383--399, 2016.

\bibitem{nozumi2017coordinated}
M.~Nozumi, F.~Nakatsu, K.~Katoh, and M.~Igarashi.
\newblock Coordinated movement of vesicles and actin bundles during nerve
  growth revealed by superresolution microscopy.
\newblock {\em Cell reports}, 18(9):2203--2216, 2017.

\bibitem{ochs2019stabilization}
F.~Ochs, G.~Karemore, E.~Miron, J.~Brown, H.~Sedlackova, M.-B. Rask, M.~Lampe,
  V.~Buckle, L.~Schermelleh, J.~Lukas, et~al.
\newblock Stabilization of chromatin topology safeguards genome integrity.
\newblock {\em Nature}, 574(7779):571--574, 2019.

\bibitem{Okhonin1987}
V.~Okhonin.
\newblock A method of examination of sample microstructure. {SU 1374922 A1}.
  {P}riority 10.04.1986. {DOI}: 10.13140/2.1.2588.1922.
\newblock 1991.

\bibitem{o2014optimized}
K.~O’Holleran and M.~Shaw.
\newblock Optimized approaches for optical sectioning and resolution
  enhancement in 2d structured illumination microscopy.
\newblock {\em Biomedical optics express}, 5(8):2580--2590, 2014.

\bibitem{phillips2020cryosim}
M.~A. Phillips, M.~Harkiolaki, D.~M.~S. Pinto, R.~M. Parton, A.~Palanca,
  M.~Garcia-Moreno, I.~Kounatidis, J.~W. Sedat, D.~I. Stuart, A.~Castello,
  et~al.
\newblock Cryosim: super-resolution 3d structured illumination cryogenic
  fluorescence microscopy for correlated ultrastructural imaging.
\newblock {\em Optica}, 7(7):802--812, 2020.

\bibitem{Prakash121061}
K.~Prakash.
\newblock Laser-free super-resolution microscopy.
\newblock {\em bioRxiv}, 2020.

\bibitem{prakash2021molecular}
K.~Prakash.
\newblock At the molecular resolution with minflux?
\newblock 2021.

\bibitem{rego2012nonlinear}
E.~H. Rego, L.~Shao, J.~J. Macklin, L.~Winoto, G.~A. Johansson,
  N.~Kamps-Hughes, M.~W. Davidson, and M.~G. Gustafsson.
\newblock Nonlinear structured-illumination microscopy with a photoswitchable
  protein reveals cellular structures at 50-nm resolution.
\newblock {\em Proceedings of the National Academy of Sciences},
  109(3):E135--E143, 2012.

\bibitem{reymond2020modulation}
L.~Reymond, T.~Huser, V.~Ruprecht, and S.~Wieser.
\newblock Modulation-enhanced localization microscopy.
\newblock {\em Journal of Physics: Photonics}, 2(4):041001, 2020.

\bibitem{reymond2019simple}
L.~Reymond, J.~Ziegler, C.~Knapp, F.-C. Wang, T.~Huser, V.~Ruprecht, and
  S.~Wieser.
\newblock Simple: Structured illumination based point localization estimator
  with enhanced precision.
\newblock {\em Optics express}, 27(17):24578--24590, 2019.

\bibitem{russell2021mmsim}
C.~T. Russell and M.~Shaw.
\newblock mmsim: An open toolbox for accessible structured illumination
  microscopy.
\newblock {\em bioRxiv}, 2021.

\bibitem{rust2006stochastic}
M.~J. Rust, M.~Bates, and X.~Zhuang.
\newblock Stochastic optical reconstruction microscopy (storm) provides
  sub-diffraction-limit image resolution.
\newblock {\em Nature methods}, 3(10):793, 2006.

\bibitem{schaefer2004structured}
L.~Schaefer, D.~Schuster, and J.~Schaffer.
\newblock Structured illumination microscopy: artefact analysis and reduction
  utilizing a parameter optimization approach.
\newblock {\em Journal of microscopy}, 216(2):165--174, 2004.

\bibitem{schermelleh2008subdiffraction}
L.~Schermelleh, P.~M. Carlton, S.~Haase, L.~Shao, L.~Winoto, P.~Kner, B.~Burke,
  M.~C. Cardoso, D.~A. Agard, M.~G. Gustafsson, et~al.
\newblock Subdiffraction multicolor imaging of the nuclear periphery with 3d
  structured illumination microscopy.
\newblock {\em Science}, 320(5881):1332--1336, 2008.

\bibitem{schermelleh2019super}
L.~Schermelleh, A.~Ferrand, T.~Huser, C.~Eggeling, M.~Sauer, O.~Biehlmaier, and
  G.~P. Drummen.
\newblock Super-resolution microscopy demystified.
\newblock {\em Nature cell biology}, 21(1):72--84, 2019.

\bibitem{schermelleh2010guide}
L.~Schermelleh, R.~Heintzmann, and H.~Leonhardt.
\newblock A guide to super-resolution fluorescence microscopy.
\newblock {\em Journal of Cell Biology}, 190(2):165--175, 2010.

\bibitem{shao2011super}
L.~Shao, P.~Kner, E.~H. Rego, and M.~G. Gustafsson.
\newblock Super-resolution 3d microscopy of live whole cells using structured
  illumination.
\newblock {\em Nature methods}, 8(12):1044--1046, 2011.

\bibitem{shaw2017creim}
M.~Shaw, A.~Bella, and M.~G. Ryadnov.
\newblock Creim: Coffee ring effect imaging model for monitoring protein
  self-assembly in situ.
\newblock {\em The journal of physical chemistry letters}, 8(19):4846--4851,
  2017.

\bibitem{shaw2015high}
M.~Shaw, L.~Zajiczek, and K.~O’Holleran.
\newblock High speed structured illumination microscopy in optically thick
  samples.
\newblock {\em Methods}, 88:11--19, 2015.

\bibitem{sheppard2017image}
C.~J. Sheppard, M.~Castello, G.~Tortarolo, G.~Vicidomini, and A.~Diaspro.
\newblock Image formation in image scanning microscopy, including the case of
  two-photon excitation.
\newblock {\em JOSA A}, 34(8):1339--1350, 2017.

\bibitem{sir2011primary}
J.-H. Sir, A.~R. Barr, A.~K. Nicholas, O.~P. Carvalho, M.~Khurshid, A.~Sossick,
  S.~Reichelt, C.~D'Santos, C.~G. Woods, and F.~Gergely.
\newblock A primary microcephaly protein complex forms a ring around parental
  centrioles.
\newblock {\em Nature genetics}, 43(11):1147--1153, 2011.

\bibitem{smeets2014three}
D.~Smeets, Y.~Markaki, V.~J. Schmid, F.~Kraus, A.~Tattermusch, A.~Cerase,
  M.~Sterr, S.~Fiedler, J.~Demmerle, J.~Popken, et~al.
\newblock Three-dimensional super-resolution microscopy of the inactive x
  chromosome territory reveals a collapse of its active nuclear compartment
  harboring distinct xist rna foci.
\newblock {\em Epigenetics \& chromatin}, 7(1):1--27, 2014.

\bibitem{song2016fast}
L.~Song, H.-W. Lu-Walther, R.~F{\"o}rster, A.~Jost, M.~Kielhorn, J.~Zhou, and
  R.~Heintzmann.
\newblock Fast structured illumination microscopy using rolling shutter
  cameras.
\newblock {\em Measurement Science and Technology}, 27(5):055401, 2016.

\bibitem{sonnen20123d}
K.~F. Sonnen, L.~Schermelleh, H.~Leonhardt, and E.~A. Nigg.
\newblock 3d-structured illumination microscopy provides novel insight into
  architecture of human centrosomes.
\newblock {\em Biology open}, 1(10):965--976, 2012.

\bibitem{wang2021ucsim2}
H.~Wang, R.~Lachmann, B.~Marsikova, R.~Heinzmann, and B.~Diederich.
\newblock Ucsim2: 2d structured illumination microscopy using uc2.
\newblock {\em bioRxiv}, pages 2021--01, 2021.

\bibitem{wu2018faster}
Y.~Wu and H.~Shroff.
\newblock Faster, sharper, and deeper: structured illumination microscopy for
  biological imaging.
\newblock {\em Nature methods}, 15(12):1011--1019, 2018.

\bibitem{yeh2019computational}
L.-H. Yeh, S.~Chowdhury, and L.~Waller.
\newblock Computational structured illumination for high-content fluorescence
  and phase microscopy.
\newblock {\em Biomedical optics express}, 10(4):1978--1998, 2019.

\bibitem{york2013instant}
A.~G. York, P.~Chandris, D.~Dalle~Nogare, J.~Head, P.~Wawrzusin, R.~S. Fischer,
  A.~Chitnis, and H.~Shroff.
\newblock Instant super-resolution imaging in live cells and embryos via analog
  image processing.
\newblock {\em Nature methods}, 10(11):1122--1126, 2013.

\bibitem{york2012resolution}
A.~G. York, S.~H. Parekh, D.~Dalle~Nogare, R.~S. Fischer, K.~Temprine,
  M.~Mione, A.~B. Chitnis, C.~A. Combs, and H.~Shroff.
\newblock Resolution doubling in live, multicellular organisms via multifocal
  structured illumination microscopy.
\newblock {\em Nature methods}, 9(7):749--754, 2012.

\end{thebibliography}

\bibliographystyle{abbrv}

\addtocounter{figure}{-1}
\begin{figure} [t!]
  \caption{\textbf{Biological super-resolution imaging with 3D-SIM.} (A) Mouse C127 mammary epithelial cell nucleus replication labelled with 5-ethenyl-2'-deoxyuridine (EdU, red) for 15 min before fixation with formaldehyde. The thymidine analogue EdU is incorporated into newly synthesized DNA of S-phase cells (here mid-to-late S phase) and detected via click-chemistry with Alexa Fluor 594 azide. DNA is labelled with 4', 6-diamidino-2-phenylindole (DAPI, cyan). Single z-section of an image stack is shown with conventional wide-field illumination (top, left), structured illumination (1 of 15 raw images acquired per z-plane with laterally shifted and rotated stripes; bottom, left), and after 3D-SIM reconstruction (right). Note that 3D-SIM resolves higher-order domain organisation of chromatin and DNA-free interchromatin regions (inset), as well as the location of nuclear pores in the peripheral chromatin layer visible as DAPI void dots in the central region of the nucleus. Scale bar: 5 µm and 1 µm (inset). (B) Corresponding frequency distribution of the DAPI signal in Fourier (reciprocal) space. Concentric rings indicate the respective spatial resolution in µm. Spots in the raw SI frequency plot (arrowheads) correspond to first and second order stripes in the image (generated by a three-beam interference; only coarse first order stripes are visible in the image). (C) Orthogonal cross section and corresponding frequency distribution of the same dataset. The arrowheads indicate the position of the z-section shown in (A). Note the 2-fold extended frequency distribution in the reconstructed data in both lateral and axial direction, that includes the filling-in of the missing frequencies along the z-axis of the wide-field frequency plot.}
\end{figure}

\end{document}